\documentclass[a4paper,english]{llncs}
\usepackage[T1]{fontenc}
\usepackage[latin9]{inputenc}
\usepackage{amsmath}
\usepackage{amssymb}
\usepackage{graphicx}

\makeatletter

\pdfpageheight\paperheight
\pdfpagewidth\paperwidth

\numberwithin{equation}{section}
\numberwithin{figure}{section}

\date{}

\usepackage{tikz}
\usetikzlibrary{shapes,arrows,patterns,mindmap,trees,positioning,fit,calc}

\makeatother

\usepackage{babel}
\begin{document}

\title{Dining Cryptographers with 0.924 Verifiable Collision Resolution}

\author{Christian Franck}

\institute{~}
\maketitle
\begin{abstract}
The dining cryptographers protocol implements a multiple access channel
in which senders and recipients are anonymous. A problem is that a
malicious participant can disrupt communication by deliberately creating
collisions. We propose a computationally secure dining cryptographers
protocol with collision resolution that achieves a maximum stable
throughput of 0.924 messages per round and which allows to easily
detect disruptors.
\end{abstract}

\section{Introduction}

Protocols for untraceable communication have received much attention
recently as they can help us protect our privacy and avoid cyber espionage.
The aim of these protocols is not to encrypt messages but to prevent
an attacker from determining who is communicating with whom. 

The dining cryptographers protocol \cite{chaum1988dcp} is the most
secure protocol for untraceable communication known in computer science.
This multi-party protocol implements a multiple access channel in
which senders and recipients of messages remain anonymous. Unlike
other primitives like mixes \cite{chaum1981uem} and onion routing
\cite{goldschlag1996hri,dingledine2004tsg}, it does not require a
trusted third party and it is not vulnerable to network based attacks
like traffic shaping. 

The problem is that messages collide when multiple senders attempt
to transmit a message at the same time. Even worse, malicious participants
can disrupt the communication by deliberately creating collisions
all the time. Such disruptors are hard to identify because the anonymity
of the honest senders must be preserved. 

Recent computationally secure variants of the dining cryptographers
protocol use an anonymous reservation phase in order to avoid collisions
and a technique based on zero-knowledge proofs to detect disruptors
\cite{FranckMscThesis,corrigan2010dissent,corrigan2013proactively}.
However, the implementation of such an anonymous reservation phase
is complicated and reservations do not adapt well to situations where
participants are frequently joining or leaving the group.

The present paper shows that one can address collisions as they occur
using a collision resolution algorithm and still prevent disruption
by a malicious participant. First, we show that with a modified SICTA
collision resolution algorithm a maximum stable throughput (MST) of
0.924 packets per round can be achieved for the dining cryptographers
protocol. Then, we show that it is possible to use zero-knowledge
proofs to verify that each participant properly executes the collision
resolution algorithm.

Compared to existing techniques our approach is easier to implement
as there is no a reservation phase. Further, it adapts better to situation
where participants are joining and leaving. We see possible applications
in the fields of electronic voting and low latency anonymous communication.

The rest of the paper is organized as follows. Section 2 contains
preliminaries and definitions. In section 3, we discuss collision
resolution with SICTA. In section 4 we show how disruptors can be
detected. In section 5 we discuss related work, in section 6 we present
possible applications, and we conclude in section~7.

\section{Preliminaries }

In this section, we briefly review the principle behind the dining
cryptographers protocol and a technique to implement it efficiently.

\subsection*{Dining Cryptographers}

In one round of the dining cryptographers protocol \cite{chaum1988dcp},
every participant broadcasts a ciphertext ($O$), which may or may
not contain a message ($M$). (To keep the description simple, we
assume that the participants have reliable broadcast channels at their
disposal.) The encryption vanishes when the ciphertexts of all participants
are combined (e.g., $C:=\prod_{i}O^{(i)}$). If exactly one ciphertext
contains a message, then this message appears (e.g., $C=M$). However,
there is a collision when several ciphertexts contain a message (e.g.,
$C=M\cdot M'\cdot M''$). We assume that messages are encoded with
a checksum, so that it is possible to distinguish between a message
and a collision of messages.

\subsection*{Generation of Ciphertexts}

We generate ciphertexts as described by Golle and Juels in~\cite{golle2004dcr}.
The advantage of this technique, which is based on the Diffie-Hellman
key agreement, is that after a single setup phase the participants
can generate ciphertexts for a large number of rounds. To this effect,
participants share finite groups $H=\langle h\rangle$ and $G=\langle g\rangle$
and a bilinear map $e:H\times H\rightarrow G$ such that $e(h^{a},h^{b})=e(h,h)^{ab}=g^{ab}$
for $a,b\in\mathbb{Z}$. The Bilinear Decisional Diffie-Hellman (BDDH)
problem is assumed to be hard in $H$ and $G$. Each participant has
a private key $x$ and the corresponding public keys $\bar{y}=h^{x}$
and $y=g^{x}$ are known to all participants.

In a round~$j$ of the protocol, each participant then generates
a ciphertext $O_{j}\in G$ which has an algebraic structure. This
means that $O_{j}$ is either of the form
\[
O_{j}=A_{j}{}^{x}
\]
or of the form
\[
O_{j}=A_{j}{}^{x}M,
\]
wherein $x\in\mathbb{Z}$ is a secret key and $M\in G$ is a message.
As we have the BDDH assumption, one cannot distinguish whether $O_{j}$
contains a message $M$ or not. 

The value $A_{j}$ is public and it is based on the public keys $\bar{y}=h^{x}$
of the other participants and on a public random value $R_{j}$ which
is different in every round. For example, $n$ participants $P^{(1)},...,P^{(n)}$
compute 
\[
A_{j}^{(i)}=e\left(\prod_{k=1}^{i-1}\bar{y}^{(k)}\prod_{k=i+1}^{n}1/\bar{y}^{(k)},R_{j}\right),
\]
The so obtained values $A_{j}^{(i)}$ are different in each round
and have the property that they cancel when they are multiplied. I.e.,
\[
\prod_{k=1}^{n}\left(A_{j}^{(k)}\right)^{x^{(k)}}=1.
\]
Therefore, only messages remain when a recipient multiplies the ciphertexts
$O_{j}^{(1)}...O_{j}^{(n)}$ provided by all the participants.

\section{Collision Resolution with SICTA}

In this section we explain the SICTA algorithm (Successive Inference
Cancellation Tree Algorithm) \cite{yu2005sicta} and show that in
the context of the dining cryptographers protocol we can reach throughput
to 0.924 messages per round.

\subsection*{Collision Resolution}

\begin{figure}[t]
\begin{centering}
\begin{tikzpicture}[yscale=0.8]
\begin{scope}[level distance=1.0cm,
sibling distance=4cm,level/.style={sibling distance=3.5cm/#1},
block/.style ={rectangle, draw=black, thick, text centered,  inner sep=0.15cm,font=\small},
blockx/.style ={rectangle, dashed, draw=black, very thick, text centered,  inner sep=0.15cm,font=\small}]
\node[block] (n1){$M_1M_2M_3M_4M_5$}
child {node (n21)[block] {$M_2M_4$}
child {node (n2x)[block,yshift=-0.8cm] {$M_2$}}
child {node (n2a)[blockx,yshift=-1.6cm] {$M_4$}}}
child {node (n22)[blockx,yshift=-0.8cm] {$M_1M_3M_5$}
child {node (n31)[block,yshift=-1.6cm] {$M_3$}}
child {node (n32)[blockx,yshift=-2.4cm] {$M_1M_5$}
child {node (n3f)[block] {$M_1$}}
child {node (n3g)[blockx,yshift=-0.8cm] {$M_5$}}}
};



\path []($(n1)+ (-5.0cm,0.5cm)$) -- node[above]{$C_j$} +(10cm,0);

\draw [densely dotted]($(n1)+ (-5.0cm,0.5cm)$) node [above,xshift=0.8cm]{round id $j$}-- +(12.1cm,0);
\draw [densely dotted]($(n1)+ (-5.0cm,-0.5cm)$) node [above,xshift=0.8cm]{$1$}-- +(12.1cm,0) node[above,anchor=south west,xshift=-2.5cm]{$C_1=\prod^n_{i=1}O^{(i)}_1$};
\draw [densely dotted]($(n1)+ (-5.0cm,-1.5cm)$) node [above,xshift=0.8cm]{$2$}-- +(12.1cm,0) node[above,anchor=south west,xshift=-2.5cm]{$C_2=\prod^n_{i=1}O^{(i)}_2$};
\draw [densely dotted]($(n1)+ (-5.0cm,-2.5cm)$) node [above,xshift=0.8cm]{$3$}-- +(12.1cm,0) node[above,anchor=south west,xshift=-2.5cm]{$C_3=C_1/C_2$};
\draw [densely dotted]($(n1)+ (-5.0cm,-3.5cm)$) node [above,xshift=0.8cm]{$4$}-- +(12.1cm,0) node[above,anchor=south west,xshift=-2.5cm]{$C_4=\prod^n_{i=1}O^{(i)}_4$};
\draw [densely dotted]($(n1)+ (-5.0cm,-4.5cm)$) node [above,xshift=0.8cm]{$5$}-- +(12.1cm,0) node[above,anchor=south west,xshift=-2.5cm]{$C_5=C_2/C_4$};
\draw [densely dotted]($(n1)+ (-5.0cm,-5.5cm)$) node [above,xshift=0.8cm]{$6$}-- +(12.1cm,0) node[above,anchor=south west,xshift=-2.5cm]{$C_6=\prod^n_{i=1}O^{(i)}_6$};
\draw [densely dotted]($(n1)+ (-5.0cm,-6.5cm)$) node [above,xshift=0.8cm]{$7$}-- +(12.1cm,0) node[above,anchor=south west,xshift=-2.5cm]{$C_7=C_3/C_6$};
\draw [densely dotted]($(n1)+ (-5.0cm,-7.5cm)$) node [above,xshift=0.8cm]{$14$}-- +(12.1cm,0) node[above,anchor=south west,xshift=-2.5cm]{$C_{14}=\prod^n_{i=1}O^{(i)}_{14}$};
\draw [densely dotted]($(n1)+ (-5.0cm,-8.5cm)$) node [above,xshift=0.8cm]{$15$}-- +(12.1cm,0) node[above,anchor=south west,xshift=-2.5cm]{$C_{15}=C_{7}/C_{14}$};

\node[inner sep=0cm,fit=(n1) (n3g) (n2x)] (la) {};
\end{scope}

\end{tikzpicture}
\par\end{centering}

\caption{Exemplary binary collision resolution tree with successive inference
cancellation (SICTA). In rounds 1,2,4,6 and 14, ciphertexts $O_{j}$
are transmitted, and $C_{j}$ is computed using these ciphertexts.
In rounds 3,5,7 and 15, no data is transmitted and $C_{j}$ is computed
using data from the parent and the sibling node.}
\label{sictafig}
\end{figure}
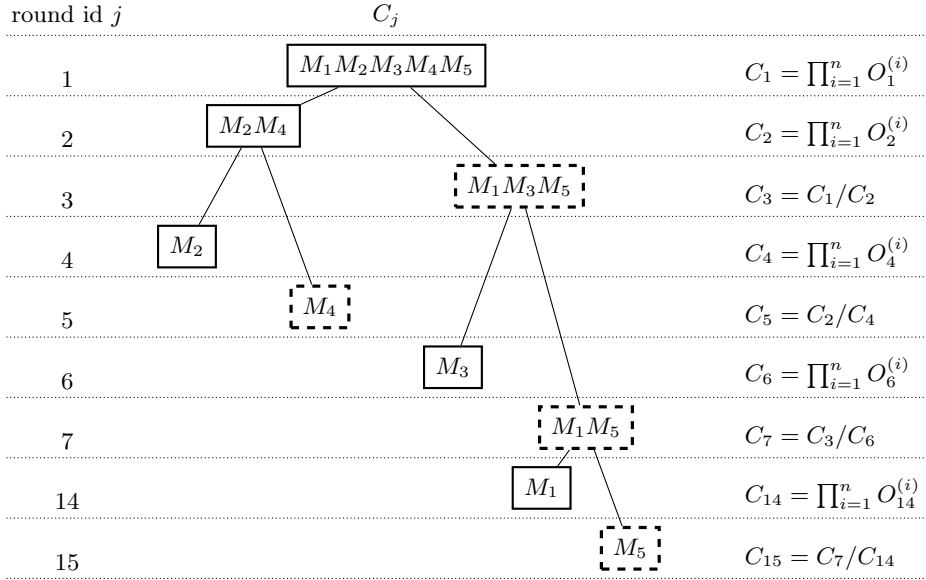

Assume in a round~$j$ each participant provides a ciphertext $O_{j}$.
If several of these ciphertexts contain a message, the combination
of all these ciphertexts $C_{j}:=\prod_{i=1}^{n}O_{j}^{(i)}$ only
provides a multiplication of all the messages and no meaningful information
is transmitted. The purpose of a collision resolution algorithm is
to resolve such a collision by resending the involved messages in
later rounds. 

SICTA is a binary tree algorithm, in which a collision of messages
is repeatedly split until all messages have been transmitted. When
there is a collision in one round, two subsequent rounds are dedicated
to the resolution of this collision. Each message involved involved
in the collision is then retransmitted at random in one of two dedicated
rounds. This process is repeated recursively until all collisions
are resolved. An example of a SICTA collision resolution tree is shown
in Figure~\ref{sictafig}. To simplify the description we adapt our
notation to binary trees; when a collision that occurs in round $j$
we assume that the rounds $2j$ and $2j+1$ are dedicated for the
resolution. SICTA uses a technique called inference cancellation to
reduce the number of transmissions. As we have $C_{j}=C_{2j}\cdot C_{2j+1}$,
it is not necessary to transfer any $O_{2j+1}$ for round $2j+1$.
The value $C_{2j+1}$ can be inferred from $C_{j}$ and $C_{2j}$
by computing $C_{2j+1}=C_{j}/C_{2j}$. For this inference cancellation
to work, the algorithm operates in blocked access mode, which means
that no new message may be sent until all collisions are resolved.

\subsection*{Performance }

Let us consider the maximum stable throughput (MST), which denotes
the maximal input rate (messages/round) for which all messages have
a finite delay. Therefore we define $S_{k}$ as the average number
of rounds needed to resolve a collision of k messages, and we consider
the throughput $k/S_{k}$. 

\begin{figure}[t]
\begin{centering}
\includegraphics{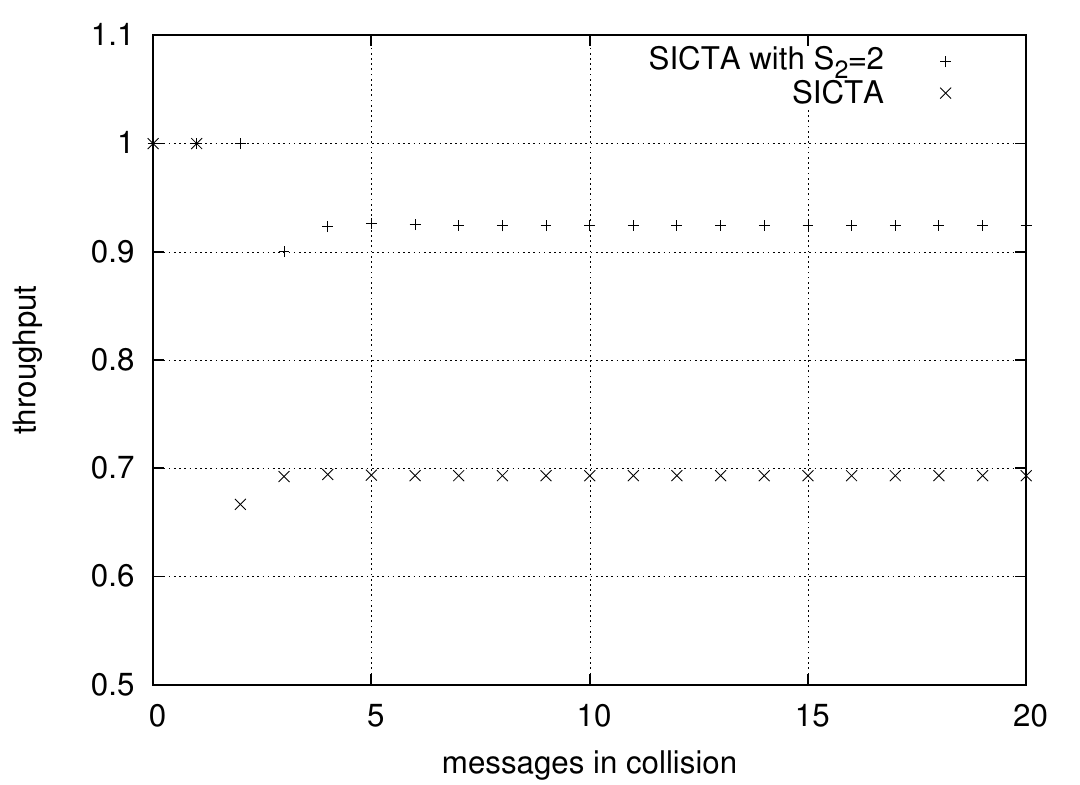}
\par\end{centering}

\caption{Performance of collision resolution with SICTA.}
\label{Figure_Throughput}
\end{figure}

A collision of $k$ messages is split into two collisions with $i$
and $k-i$ messages with a probability ${k \choose i}2^{-k}$. Thus
we have
\[
S_{k}=\sum_{i=0}^{k}{k \choose i}2^{-k}(S_{i}+S_{k-i}).
\]
With ${k \choose i}={k \choose k-i}$ this can be written as
\[
S_{k}=\sum_{i=0}^{k}{k \choose i}2^{1-k}S_{i}
\]
and after removing the recursion we obtain
\[
S_{k}=\frac{2^{1-k}}{1-2^{1-k}}\sum_{i=0}^{k-1}{k \choose i}S_{i}.
\]
As 'collisions' with $0$ or $1$ messages take only 1 round, we
have $S_{0}=S_{1}=1$. The throughput $k/S_{k}$ for increasing values
of $k$ is shown in Figure~\ref{Figure_Throughput}. We observe for
SICTA the known MST of 0.693. 

We can achieve a higher throughput by exploiting the fact that in
the dining cryptographers protocol all senders are also receivers.
After a collision of two messages, the two respective senders can
recover each other's message by removing their own from the collision.
They can then avoid a further collision by using a rule that for instance
only the numerically smaller message is resent. This way, collisions
of two messages are always resolved in two rounds. I.e., we have $S_{2}=2$,
which leads to a MST of 0.924.

So we  have just computed the possible throughput of the channel and
seen that efficient collision resolution is possible. We have done
this under the assumption that every participant is honest and that
no disruption takes place. This assumption is reasonable, as we show
in the next section that disruptors can easily be detected and eliminated
from the group. Being exceptional events, disruptions have no impact
on the asymptotic ($\mbox{number of rounds}\rightarrow\infty$) behavior
of the channel. 

\section{Detecting Disruptors}

In this section, we show that disruptors are easy to detect. We first
present techniques using zero-knowledge proofs to prove statements
about the retransmission of messages, and then we show how these techniques
can be used to verify that each participant correctly performs the
SICTA algorithm.

\subsection*{Zero-Knowledge Proofs for the Retransmission of Messages}

It was shown in \cite{golle2004dcr} that the algebraic structure
of the ciphertexts makes it possible to prove statements about them
using zero-knowledge proofs. Such a zero-knowledge proof allows a
prover to prove to a verifier that a given statement holds, without
giving the verifier any further information. I.e., the verifier cannot
compute anything that he could not have computed before. One can for
instance prove the equality of discrete logarithms to different bases,
and logical $\wedge$ (and) and $\vee$ (or) combinations of such
statements~\cite{camenisch1997psg}. It is also possible to prove
the inequality of logarithm to different bases \cite{camenisch2003practical}.

Existing zero-knowledge proofs used in dining cryptographers protocols
contain statements about individual ciphertexts. E.g., the statement
\[
\log_{A_{1}}(O_{1})=\log_{g}y
\]
holds when ciphertext $O_{1}$ is empty (i.e., $O_{1}=A_{1}^{x}$).
As a reminder, $x$ is a secret key of the participant and $y=g^{x}$
the corresponding public key.

To verify the correct execution of the SICTA collision resolution
protocols we use a new kind of statements, which hold when there is
a relation between two or more ciphertexts coming from the same participant.
E.g., the statement
\[
\log_{A_{1}/A_{2}}(O_{1}/O_{2})=\log_{g}y
\]
holds when both ciphertexts $O_{1}$ and $O_{2}$ encode the same
message $M$ (or when both encode no message). It is thus possible
to construct more complex statements in order to verify the retransmission
of a message.

\begin{example}
The ciphertext $O_{2}$ either contains no message, or the same message
as~$O_{1}$, when the statement
\[
\left(\log_{A_{2}/A_{1}}(O_{2}/O_{1})=\log_{g}y\right)\vee\left(\log_{A_{2}}(O_{2})=\log_{g}y\right)
\]
holds.
\begin{example}
\label{example_2}At most one ciphertext out of $O_{2},...,O_{k}$
contains the same message as $O_{1}$ (and the rest of $O_{2},...,O_{k}$
contain no message), when the statement
\[
\left(\log_{A_{2}...A_{j}/A_{1}}(O_{2}...O_{j}/O_{1})=\log_{g}y\right)\vee\left(\log_{A_{j}}(O_{j})=\log_{g}y\right)
\]
holds for $j\in\{2,...,k\}$. (Note that it is not sufficient to consider
only the last statement, as a participant could encode $O_{2}=A_{2}^{x}E$
and $O_{3}=A_{3}^{x}E^{-1}$ instead of $O_{2}=A_{2}^{x}$ and $O_{3}=A_{3}^{x}$.
In the multiplication $O_{2}O_{3}...$, the factors $E$ and $E^{-1}$
would cancel, and the statement would hold. It is therefore necessary
to consider each statement for $j\in\{2,...,k\}$.) 
\end{example}
\end{example}

\subsection*{Verification of Standard SICTA with a MST of 0.693}

We now show how the techniques from the previous section can be used
by the participants to prove that they correctly executed the collision
resolution algorithm, without revealing if they are sending a message
or not (so that the senders of the messages remain anonymous).
\begin{figure}[t]
\begin{centering}
\begin{tikzpicture}[level distance=1.0cm,
sibling distance=5cm,level/.style={sibling distance=1.2cm/#1},
block/.style ={rectangle, draw=black, thick, text centered,  inner sep=0.15cm,text height=1.8ex,minimum width=5ex,font=\small},
blockx/.style ={rectangle, dashed,draw=black, thick, text centered,  inner sep=0.15cm,text height=1.8ex,minimum width=5ex,font=\small}]

\begin{scope}
\node[block] (a1){}
child {node (a2)[block] {}}
child {node (a3)[blockx,yshift=-1cm] {}};
\end{scope}

\begin{scope}[xshift=3.5cm]
\node[block] (b1){$M$}
child {node (b2)[block] {$M$}}
child {node (b3)[blockx,yshift=-1cm] {}};
\end{scope}

\begin{scope}[xshift=7cm]
\node[block] (c1){$M$}
child {node (c2)[block] {}}
child {node (c3)[blockx,yshift=-1cm] {$M$}};
\end{scope}

\node[inner sep=0cm,fit=(a1) (a2) (a3)] (la) {};
\node at (la.south)[below=0.3cm,inner sep=0,font=\small, text width=3.3cm, text centered] {(a) No Retransmission in round $2j$.};
\node[inner sep=0cm,fit=(b1) (b2) (b3)] (lb) {};
\node at (lb.south)[below=0.3cm,inner sep=0,font=\small, text width=3.3cm, text centered] {(b) Retransmission  in round $2j$.};
\node[inner sep=0cm,fit=(c1) (c2) (c3)] (lc) {};
\node at (lc.south)[below=0.3cm,inner sep=0,font=\small, text width=3.3cm, text centered] {(c) No Retransmission  in round $2j$.};

\draw [densely dotted]($(a1)+ (-3.0cm,0.5cm)$) node [above,xshift=0.7cm]{round id}-- +(12.1cm,0);
\draw [densely dotted]($(a1)+ (-3.0cm,-0.5cm)$) node [above,xshift=0.7cm]{$j$}-- +(12.1cm,0);
\draw [densely dotted]($(a1)+ (-3.0cm,-1.5cm)$) node [above,xshift=0.7cm]{$2j$}-- +(12.1cm,0);
\draw [densely dotted]($(a1)+ (-3.0cm,-2.5cm)$) node [above,xshift=0.7cm]{$2j+1$}-- +(12.1cm,0);

\end{tikzpicture}

\par\end{centering}

\caption{Collision resolution in SICTA. Only a message involved in a collision
in round $j$ may be retransmitted either in round $2j$. The round
$2j+1$ is virtual; no transmission takes place. No new message may
be sent until all collisions are resolved.}
\label{figure1}
\end{figure}
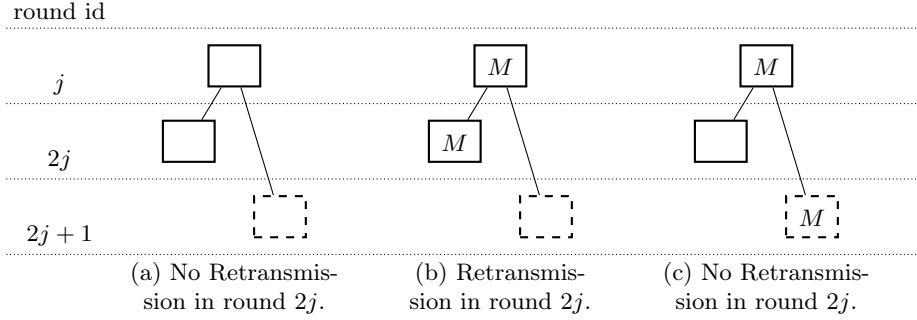

Correct participation in the standard SICTA algorithm means that a
participant may only retransmit a message in round $2j$ if he already
transmitted that message in round $j$. Remember that SICTA operates
in blocking mode an that no new message may be sent until the resolution
has finished. This principle, which is illustrated in Figure~\ref{figure1},
means that
\begin{itemize}
\item if a participant transmits $O_{j}=A_{j}^{x}$ in round $j$, then
he must transmit $O_{2j}=A_{2j}^{x}$ in round $2j$; and
\item if a participant transmits $O_{j}=A_{j}^{x}M$ in round $j$, then
he must transmit either $O_{2j}=A_{2j}^{x}$ or $O_{2j}=A_{2j}^{x}M$
in round $2j$. \\
(Note that sending no message equals to sending $M=1$. If the participant
did not send a message in round $j$, this means that $O_{j}=A_{j}^{x}\cdot1$.
The participant then has the 'choice' between retransmitting $O_{2j}=A_{2j}^{x}$
or $O_{2j}=A_{2j}^{x}\cdot1$ in round $2j$. I.e., he is forced to
retransmit $O_{2j}=A_{2j}^{x}\cdot1=A_{2j}^{x}$ and may not send
a message.)
\end{itemize}

Using the techniques from the previous section, each participant can
prove that his ciphertext $O_{2j}$ is correct, without revealing
if whether it contains a message or not. To do this, the participant
generates a zero-knowledge proof that proves that 
\begin{equation}
\left(\log_{A_{j}/A_{2j}}(O_{j}/O_{2j})=\log_{g}y\right)\vee\left(\log_{A_{2j}}(O_{2j})=\log_{g}y\right)\label{eq:1}
\end{equation}
holds. With this proof he can convince a verifier that he participated
correctly, without compromising the anonymity of the protocol. 

As described before, SICTA is a recursive algorithm and there are
virtual rounds during which $C_{j}$ is inferred, but no corresponding
$O_{j}$ is transmitted. It is then not possible to prove statement
(\ref{eq:1}), but luckily it is still possible to prove that $O_{2j}$
is correct. To do this, the participant proves that a message contained
in the nearest transmitted parent round was transmitted at most once
in all the branches down to $O_{2j}$. Akin to Example \ref{example_2},
a participant proves that 
\[
\left(\log_{A_{j_{t}/2}/A_{j_{1}}...A_{j_{t}}}(O_{j_{t}/2}/O_{j_{1}}...O_{j_{t}})\mbox{=}\log_{g}y\right)\vee\left(\log_{A_{2j}}(O_{2j})\mbox{=}\log_{g}y\right)
\]
holds, wherein $j_{1}:=2j$, $j_{k}:=(j_{k-1}/2)-1$ and $t$ such
that $j_{t}/2$ is the index of the nearest transmitted parent round
of round $j$.
\begin{example}
In the collision resolution process shown in Figure \ref{sictafig},
each participant shows for $O_{2}$ that 
\[
\left(\log_{A_{1}/A_{2}}(O_{1}/O_{2})=\log_{g}y\right)\vee\left(\log_{A_{2}}(O_{2})=\log_{g}y\right)
\]
holds, then for $O_{4}$ that
\[
\left(\log_{A_{2}/A_{4}}(O_{2}/O_{4})=\log_{g}y\right)\vee\left(\log_{A_{4}}(O_{4})=\log_{g}y\right)
\]
holds, then for $O_{6}$ that
\[
\left(\log_{A_{1}/A_{2}A_{6}}(O_{1}/O_{2}O_{6})=\log_{g}y\right)\vee\left(\log_{A_{6}}(O_{6})=\log_{g}y\right)
\]
holds, then for $O_{14}$ that
\[
\left(\log_{A_{1}/A_{2}A_{6}A_{14}}(O_{1}/O_{2}O_{6}O_{14})=\log_{g}y\right)\vee\left(\log_{A_{14}}(O_{14})=\log_{g}y\right)
\]
holds.

So we have shown that a participant can prove in zero-knowledge that
he properly participates in the standard SICTA collision resolution
algorithm. Any participant who is not able to prove that his output
is correct can be excluded from the group. The corresponding round
is lost and must be repeated by the remaining participants. 
\end{example}

\subsection*{Additional Verification for Optimized SICTA with a MST of 0.924}

The special technique we described in section to increase the MST
from 0.693 to 0.924 uses a deterministic rule for the retransmission
after a collision of two messages. E.g. only the message with the
lower value must be retransmitted. So we need additional verification
to detect participants that do not respect this rule. We cannot verify
this with a single zero-knowledge proof, but we can for instance use
the following approach.

If one of the two participants involved in the collision does not
respect this rule, the other one can switch back to random retransmission
in order to split the collision. Once the collision has been split
and the two messages are out, everybody sees that there must have
been a problem in a previous round, and an investigation can be started.
The message $M$ of the cheating participant is now known, and every
participant must then for instance send a zero-knowledge proof that
he did not send this message during the initial collision round (i.e.
prove that $\log_{A_{1}}O_{1}/M\ne\log_{g}y$). The disruptor will
not be able to come up with an appropriate proof and can be eliminated
from the group.

\subsection*{Further Minor Security Considerations}

Malicious participants may attempt to delay the collision resolution
process or to prevent it from terminating. For instance,
\begin{itemize}
\item colluding participants can always chose the same round to retransmit
their messages, or
\item a malicious participant can wait until all other participants have
transmitted and then choose to retransmit his message so that a collision
occurs, or 
\item a malicious participant may not send a valid message in the first
place. 
\end{itemize}
However, such malicious behavior is easy to detect. In the previously
described SICTA algorithm with a MST of 0.924, the probability that
a collision does not split is less than or equal to $1/4$ (it is
exactly $1/4$ for collisions with 3 messages). Thus, the probability
that a collision does not split $k$ times in a row is less than or
equal to $1/4^{k}$. E.g., the probability that a collision does not
split 5 times in a row is below $0.1\%$. When such malicious activity
is detected, one can require commitment before transmission and one
can use zero-knowledge proofs similar to the ones proposed in \cite{golle2004dcr}
to detect participants that are frequently involved in non-splitting
collisions. If a lower throughput is acceptable, one can go for a
simpler approach and just skip the branches of the resolution tree
that do not split after several attempts, without trying to detect
the malicious participants.

\section{Related Work}

Superposed receiving \cite{pfitzmann1987iiw,waidner1990usa} is a
collision resolution technique for the dining cryptographers protocol
that achieves throughput of 100\%. Therein, messages are elements
of an additive group. When a collision occurs, the average of the
messages values is computed and only messages whose value is less
than this average are retransmitted. Like in SICTA, inference cancellation
is used, which leads to the $100\%$ throughput. However, this approach
requires the use of an additive finite group and it cannot be implemented
using the algebraic ciphertexts that we need for efficient ciphertexts
generation and for zero-knowledge proofs.

A fully verifiable dining cryptographers protocol was proposed in~\cite{FranckMscThesis}
and rediscovered in \cite{corrigan2013proactively}. In this protocol,
we have 100\% throughput. However, there is the need for a reservation
phase which can be lengthy and cumbersome. Current systems are using
mixnets to perform the reservations and therefore they are inefficient
when only a few reservations are made. Further, they do not easily
adapt to situations where participants join or leave frequently.

\section{Applications}

Our protocol can be used to implement computationally secure anonymous
communication channels with a low latency. Another application is
the realization of secret shuffle algorithms (e.g. \cite{neff2001verifiable}).
A secret shuffle algorithm is used to obtain a shuffled list of values
from a plurality of participants, while keeping it secret which value
is coming from which participant. Existing solutions typically require
each participant to submit a value. The protocol proposed herein also
works efficiently if only a few participants have a value to submit.
In particular it may be used to shuffle anonymous public keys for
verifiable dining cryptographers protocols in which rounds are reserved
\cite{FranckMscThesis,corrigan2013proactively}.

\section{Concluding Remarks}

The main problems of the dining cryptographers protocol are collisions
and malicious participants disrupting the communication.

We have shown that with a collision resolution algorithm it is possible
to achieve a maximum stable throughput of up to 0.924 messages per
round. Further, we have shown that if we use ciphertexts with an algebraic
structure as proposed in \cite{golle2004dcr}, we can verify in zero-knowledge
that each participant properly retransmits his message during the
collision resolution process. 

Compared to other dining cryptographer protocols, our approach does
not need a reservation phase to avoid collisions. It is therefore
easier to implement and it adapts more naturally to situations where
participants are frequently joining and leaving the group. 

We see possible applications in the fields of low-latency anonymous
communication and secret shuffling.

\bibliographystyle{plain}
\bibliography{references}

\begin{thebibliography}{10}

\bibitem{camenisch1997psg}
J.~Camenisch and M.~Stadler.
\newblock {Proof systems for general statements about discrete logarithms}.
\newblock {\em Technical Report TR 260, Institute for Theoretical Computer
  Science, ETH Zurich}, Mar. 1997.

\bibitem{camenisch2003practical}
Jan Camenisch and Victor Shoup.
\newblock Practical verifiable encryption and decryption of discrete
  logarithms.
\newblock In {\em Advances in Cryptology-CRYPTO 2003}, pages 126--144.
  Springer, 2003.

\bibitem{chaum1988dcp}
D.~Chaum.
\newblock {The dining cryptographers problem: Unconditional sender and
  recipient untraceability}.
\newblock {\em Journal of Cryptology}, 1(1):65--75, 1988.

\bibitem{chaum1981uem}
D.L. Chaum.
\newblock {Untraceable electronic mail, return addresses, and digital
  pseudonyms}.
\newblock {\em Communications of the ACM}, 24(2):84--88, 1981.

\bibitem{corrigan2013proactively}
H.~Corrigan-Gibbs, D.~I. Wolinsky, and B.~Ford.
\newblock Proactively accountable anonymous messaging in verdict.
\newblock In {\em USENIX Security}, 2013.

\bibitem{corrigan2010dissent}
Henry Corrigan-Gibbs and Bryan Ford.
\newblock Dissent: accountable anonymous group messaging.
\newblock In {\em Proceedings of the 17th ACM conference on Computer and
  communications security}, pages 340--350. ACM, 2010.

\bibitem{dingledine2004tsg}
R.~Dingledine, N.~Mathewson, and P.~Syverson.
\newblock {Tor: the second-generation onion router}.
\newblock {\em Proceedings of the 13th conference on USENIX Security
  Symposium-Volume 13 table of contents}, pages 21--21, 2004.

\bibitem{FranckMscThesis}
C.~Franck.
\newblock {New Directions for Dining Cryptographers}.
\newblock Master's thesis, University of Luxembourg, Luxembourg, 2008.

\bibitem{goldschlag1996hri}
D.~Goldschlag, M.~Reed, and P.~Syverson.
\newblock {Hiding Routing Information}.
\newblock {\em LECTURE NOTES IN COMPUTER SCIENCE}, pages 137--150, 1996.

\bibitem{golle2004dcr}
P.~Golle and A.~Juels.
\newblock {Dining Cryptographers Revisited}.
\newblock {\em Advances in cryptology-EUROCRYPT 2004: International Conference
  on the Theory and Applications of Cryptographic Techniques, Interlaken,
  Switzerland, May 2-6, 2004: Proceedings}, 2004.

\bibitem{neff2001verifiable}
C~Andrew Neff.
\newblock A verifiable secret shuffle and its application to e-voting.
\newblock In {\em Proceedings of the 8th ACM conference on Computer and
  Communications Security}, pages 116--125. ACM, 2001.

\bibitem{pfitzmann1987iiw}
A.~Pfitzmann.
\newblock {How to implement ISDNs without user observability -- Some remarks}.
\newblock {\em ACM SIGSAC Review}, 5(1):19--21, 1987.

\bibitem{waidner1990usa}
M.~Waidner.
\newblock {Unconditional Sender and Recipient Untraceability in spite of Active
  Attacks}.
\newblock {\em Lecture Notes in Computer Science}, 434:302, 1990.

\bibitem{yu2005sicta}
Yingqun Yu and Georgios~B Giannakis.
\newblock Sicta: a 0.693 contention tree algorithm using successive
  interference cancellation.
\newblock In {\em INFOCOM 2005. 24th Annual Joint Conference of the IEEE
  Computer and Communications Societies. Proceedings IEEE}, volume~3, pages
  1908--1916. IEEE, 2005.

\end{thebibliography}

\end{document}